# Nano plasmon polariton modes of a wedge cross section metal waveguide

**Eyal Feigenbaum and Meir Orenstein**

*EE Department, Technion, Haifa 32000, Israel*
*meiro@ee.technion.ac.il*

**Abstract:** Optical plasmon-polariton modes confined in both transverse dimensions to significantly less than a wavelength are exhibited in open waveguides structured as sharp metal wedges. The analysis reveals two distinctive modes corresponding to a localized mode on the wedge point and surface mode propagation on the abruptly bent interface. These predictions are accompanied by unique field distributions and dispersion characteristics.

**References and Links**
1. W.L. Barnes, A. Dereux, T.W. Ebbesen, "Surface plasmon subwavelength optics," Nature **424**, 824 (2003).
2. P. Berini, "Plasmon-polariton modes guided by a metal film of a finite width bounded by different dielectrics," Opt. Express **7**, 329 (2000).
3. V.A. Podolskiy, A.K. Sarychev, V.M. Shalaev, " Plasmon modes in metal nanowires and left-handed materials," J. Nonlinear Physics and Materials **11**, 65 (2002).
4. L. Dobrzynski, A.A. Maradudin, "Electrostatic Edge Modes in a Dielectric Wedge," Phys. Rev. B **6**, 3810 (1972).
5. A. Eguiluz, A.A. Maradudin, "Electrostatic edge modes along a parabolic wedge," Phys. Rev. B **14**, 5526 (1976).
6. A.D. Boardman, R. Garcia–Molina, A. Gras–Marti, E. Louis, "Electrostatic edge modes of a hyperbolic dielectric wedge: Analytical solution," Phys. Rev. B **32**, 6045 (1985).
7. D.F.P. Pile, T. Ogawa, D.K. Gramotnev, T. Okamoto, M. Haraguchi, M. Fukui, S. Matsuo, "Theoretical and experimental investigation of strongly localized plasmons on triangular metal wedges for subwavelength waveguiding," Appl. Phys. Lett. **87,** 061106 (2005).
8. I.V. Novikov, A.A. Maradudin, "Channel polaritons," Phys. Rev. B **66**, 035403 (2002).
9. S.I. Bozhevolnyi, V.S. Volkov, E. Devaux, T.W. Ebbesen, "Channel Plasmon-Polariton Guiding by Subwavelength Metal Grooves," Phys. Rev. Lett. **95**, 046802 (2005).
10. D.F.P. Pile, D.K. Gramotnev, "Channel plasmon-polariton in a triangular groove on a metal surface," Opt. Lett. **29**, 1069 (2004).
11. G.Z. Forristall, J.D. Ingram, "Evaluation of distributions useful in Kontorovich-Lebedev transform theory," SIAM J. Math. Anal. **3**, 561 (1972).
12. B. Prade, J.Y. Vinet, "Guided optical waves in fibers with negative dielectric constant," J. Light. Tech. **12**, 6 (1994).
13. H. Reather, *Surface plasmon* (Springer,Berlin,1988).
14. M.P. Nezhad, K. Tetz, Y. Fainman, "Gain assisted propagation of surface plasmon polariton on planar metallic waveguides," Opt. Express **12**, 4072 (2004).

## 1. Introduction

Sub wavelength confinement of guided waves is highly advantageous for downscaling optical components to the nano-scale, such as for inter-chip optical wiring and nano-systems interconnects. Here we present modal analysis of a metallic waveguide with a cross-section shaped as a narrow wedge, exhibiting modal size hundred times smaller than the wavelength in each transverse dimension.

Surface Plasmon-Polariton (SPP) is a guided wave confined in 1D to a sub-wavelength cross-section, though infinite in the other transverse dimension [1]. Confinement in both the transverse dimensions is achievable using the plasmonic slow waves and specific topologies, analogous to "standard" dielectric waveguides – namely metallic nano-wires, which were discussed elsewhere [2,3]. However the surface guiding of SPP allows also for open contour topologies, relaxing the waveguide transversal dimension constraint.

While a metallic surface supports 1D confined SPP, a modification in the curvature of a metallic surface, e.g., a wedge shaped metallic structure, is expected to support Plasmon-Polariton confined in 2D. The enhanced field of the guided wave in the vicinity of the wedge tip may be employed to enhance matter-field interactions. Here we analyze the wedge waveguide in closed form to understand the role of parameters (e.g. wedge angle and dielectric constants), the dispersion relations and the modal confinement.

Full fledged analytic solutions for optical modes of metallic wedge waveguides were never reported, and the most detailed analysis employed a scalar wave-equation assuming electrostatic regime for sharp [4] and smooth [5,6] topologies. Although simplifying the analysis, this eliminates the mere issue of the full vectorial nature of the hybrid modes and their structural dispersion while discarding the retardation effects. The electrostatic solution obtained in [4] is an asymptotical approximation to the one obtained here.

Recently, localized plasmons propagating on metal wedge were demonstrated using FDTD simulation and in experiments [7]. The analysis presented here provides understanding of the structure of the modes and their parametric dependence, among which proposing the existence of a critical wedge angle similar to the one reported for modes of different symmetry, and should assist in enhanced design of such novel nano-waveguides.

Worth mentioning is a related complementary problem of propagation along smooth channels carved in metal, analyzed using a modified moments method [8]. However, the basis set of the functions used there is not suitable for sharp wedge configuration and thus not appropriate for the problem at hand. Additionally, resolving the field profile in [8] involves a tedious solution of a set of integral equations. In the current work we derive closed form solutions for sharp-wedge cross-section metal waveguides, exhibiting a substantial field concentration near the wedge point and study the interesting dispersion characteristics of the guided modes. This analysis can be trivially enhanced to the complementary structures of v-grooves in metal, recently studied via simulations and experimentally [9,10].

**2. Analysis of wedge plasmonic modes**

The analysis of the metal wedge waveguide surrounded by a dielectric is performed in cylindrical coordinates system $\{r,\theta,z\}$ assuming indefinitely long wedge, both in the radial and axial directions, as shown in Fig. 1(a). Though actual wedges are finite, the plasmonic fields are located around the wedge tip, which validate this assumption.

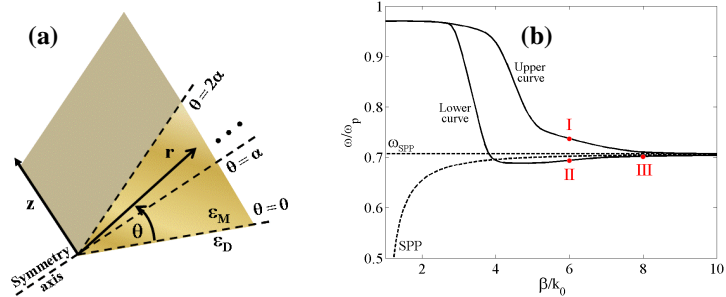

Fig. 1. Metal wedge waveguide surrounded by air. (a) schematics (b) $E^eH^o$ mode dispersion relations for $36^0$ gold wedge.

Since plasmonic modes are slow waves, they exponentially decay into both air and metal, in the radial direction as well as azimuthally. In cylindrical coordinate system the radial solutions of the wave equation are the modified Bessel functions of the second kind, denoted by $K$. Satisfying the boundary conditions at each interface point is impossible for a single Modified Bessel function at each medium, thus the solution is a series of K-functions.

The order of the K-functions may be chosen as real or imaginary, i.e. set of basis function of the form $\sim K_v(qr)\exp(iv\theta)$ or $\sim K_{iv}(qr)\exp(-v\theta)$. $q$ is the radial momentum, and $v$ denoted the K-function order. Taking the order to be imaginary is more suitable since it implies azimuthal hyperbolic dependence, rather than harmonic, which better describes the azimuthal decay of the plasmonic solutions away from the interface. Moreover, the $\{K_{iv}(qr)\}_v$ is a complete set (in contrary to $\{K_v(qr)\}_v$) with basis functions being square integrable and having vanishing power at infinity (decaying faster than $r^{-1}$).

Using this set for the solution at each media: $F(r,\theta)=\int dv\, f_v\, K_{iv}(qr)\exp(-v\theta)$, the expansion coefficients, as a function of the order, $f_v$, define the v-spectrum of the solution. The general v-spectrum is continuous on the indefinite interval, $[0,\infty)$. However the compactness of this expansion allows for a sufficient representation on a limited interval, $[0,v_{max}]$. The criteria for sufficiently high $v_{max}$ were the fulfillment of the boundary conditions to 1% accuracy. As the order of a basis function is increased, the azimuthal decay is enhanced while the radial one is decreased. The confinement of the plasmonic fields around the wedge tip is translated to effectively bounding the v-spectrum. This could be inferred also from the Kontorovich-Lebedev transform [11], having K-functions with imaginary order as basis set.

Truncating the v-spectrum components above $v_{max}$ may cause errors mostly at the interfaces. Thus, constructing and testing the solution, according to the boundary conditions, on the interface assures of the solution accuracy on the entire domain.

### 3. Modal characteristics

Applying the analysis described in the previous section, the symmetry permits for two sets of solutions: odd and even about the axis of symmetry ($\theta=\alpha$). Solving the problem in terms of tangential field components (i.e. $E_z$ and $H_z$), four independent boundary conditions apply for each r>0 on the interface $\theta=0$:

$$\begin{cases}(i) & \tilde{E}_M = \tilde{E}_D & (iii) & \partial_\theta \tilde{E}_M = -f_1 \partial_\theta \tilde{E}_D - f_2 r \partial_r \tilde{H}_D \\ (ii) & \tilde{H}_M = \tilde{H}_D & (iv) & \partial_\theta \tilde{H}_M = -f_3 r \partial_r \tilde{E}_D + f_4 \partial_\theta \tilde{H}_D\end{cases} \quad (1)$$

The ~ denotes the fields at the interfaces, M and D denote the fields in metal and dielectric respectively. $f_i$ (i=1...4) coefficients are frequency and $\beta$ dependent ($\beta$ - the modal propagation constant), given explicitly in the Appendix. From Eq. (1) it is apparent that only hybrid modes are possible. Out of the four symmetries we study in this text the solutions having the E(even)-H(odd) symmetry, exited with symmetrical electrical field.

The general solution is:

$$E_z^e = \exp(i\beta z)\begin{cases}\int dv\,\{a_v K_{iv}(q_M r)\cosh(v(\theta-\alpha))\} & \theta \in [0,2\alpha] \\ \int dv\,\{b_v K_{iv}(q_D r)\cosh(v(\theta-\alpha-\pi))\} & \theta \in [2\alpha,2\pi]\end{cases} \quad (2)$$

Where the radial momentum: $q_{D,M}=(\beta^2-k_0^2\varepsilon_{D,M})^{0.5}$, $k_0$ is the free-space wave number, and $\varepsilon$ is the relative dielectric constant. For $H_z^o$ a,b,cosh are replaced by c,d,sinh in Eq. 2, v which denotes the K-function imaginary order, is real and continuous values. Since each of the basic field functions in both sides of the interface has a different r-dependence, the boundary conditions for all r values dictate a superposition of wave solutions. This is a major complication absent in the electrostatic analysis, for which one assumes substantial $\beta$-values resulting with similar basis functions on both sides of interface (i.e. $q_M=q_D$).

Approximating the v-spectrum integral as N-sized discrete series, projecting the four boundary conditions onto the base function of the dielectric $\{K_{iv}(q_D r)\}$, using orthogonality of $K_{iv}(qr)/r$, and characteristics of the K-function integral, yields 4N algebraic equations:

$$\sum_v a_v \cosh(v\alpha) G\left(v,s;{q_M}/{q_D}\right) = b_s \cosh(s(\alpha+\pi))\frac{\pi^2}{2s\,\sinh(\pi s)} \quad (3\text{-}a)$$

$$\sum_v c_v \sinh(v\alpha) G\left(v,s;{q_M}/{q_D}\right) = d_s \sinh(s(\alpha+\pi))\frac{\pi^2}{2s\,\sinh(\pi s)} \quad (3\text{-}b)$$

$$\sum_v a_v v \sinh(v\alpha) G\left(v,s;{q_M}/{q_D}\right) = -f_1 b_s \sinh(s(\alpha+\pi))\frac{\pi^2}{2\sinh(\pi s)} - f_2 \sum_v d_v \sinh(v(\alpha+\pi)) H(v,s) \quad (3\text{-}c)$$

$$\sum_v c_v v \cosh(v\alpha) G\left(v,s; q_M/q_D\right) = -f_3 \sum_v b_v \cosh(v(\alpha+\pi)) H(v,s) + f_4 d_s \cosh(s(\alpha+\pi)) \frac{\pi^2}{2\sinh(\pi s)} \quad \text{(3-d)}$$

Expressions for G and H functions are given at the Appendix. The solution of this algebraic equation set is self-consistent for a vanishing determinant – resulting in the dispersion relation of the guided modes and the field v-spectrum, from which the field distribution is extracted.

Typically the v-spectrum peak value is observed around v=0 and v=3/α for the fields in the dielectric and metal respectively and the spectrum is rapidly decaying with v. Broadest span of v-spectrum is required for the solution on interfaces.

To elucidate the basic plasmonic mode characteristics, the metal losses are not taken into account in the followings. It can be done due to the relative smallness of loss in respect to metal dispersion and for short propagation distance, as is commonly done. We use the lossless Drude model for the metal permittivity, with plasma wavelength of $\lambda_P$=137nm. The same calculation scheme may support the case of complex permittivity, namely complex $\varepsilon_M$ and $\beta$.

The analysis is exemplified for a 36 degrees gold wedge waveguide ($2\alpha=36^0$) surrounded by air. The dispersion relations are depicted in fig. 1(b), denoted by two bold curves. Both dispersion curves are reaching asymptotically the single surface SPP frequency ($\omega_{SPP}=\omega_P/(1+\varepsilon_D)^{0.5}$) as $\beta$ tends to infinity - typical for plasmonic waveguides. The single surface SPP dispersion curve is superimposed in Fig. 1(b) (dashed). The upper branch (frequency-wise) exhibits negative group velocity, which vanishes for infinite β values. The lower dispersion branch is divided by a minimum frequency point ($\omega_{min}$=0.69$\omega_p$) into negative and positive group velocity intervals. For a frequency above the cutoff frequency ($\omega_{min}$), in the interval ($\omega_{min}<\omega<\omega_{SPP}$), two propagation constants are simultaneously supported, with opposite group velocity signs. Similar phenomenon for the same symmetry is shown, e.g., for specific set up of a metal slab [12]. The positive group index at high β values is due to the single surface SPP characteristics (power always predominantly in the dielectric), while negative group index at lower β is related to higher power in the metal due to enhanced interface coupling. Similar phenomena can be seen for a metal wire configuration [12]

The frequency of the upper branch decreases monotonically with $\beta$, accompanied by tighter localization around the wedge point. The $E_z$ field component at pt. (I) is depicted in Fig. 2, and drops off symmetrically for all transverse directions – into the metal and dielectric. At the same effective index value, the lower dispersion branch exhibits different field characteristics, as demonstrated at Fig. 3. Comparison of the $E_z$ component reveals that a main lobe is still localized on the wedge point but secondary lobes appear as well on the metal interfaces. Decreasing the modal frequency, the number of side lobes as well as the respective modal cross section increase. The distribution of the pointing vector directed along z ($S_z$) at dispersion points I and II are depicted in Fig 4(a) and 4(b), accordingly. Especially at point II, although the power propagates predominantly in the dielectric, more than 90% of the mode's intensity and power ($S_z$) are guided within a cross section of 0.01% of free-space wavelength - well under the "diffraction limit". In Fig. 5 it is apparent that also for the lower branch – the mode becomes more localized towards the wedge point as β is increased.

The distinct characteristics of the two dispersion branches are associated with the 'dual role' that the wedge plays. The one dimensional discontinuity formed at the wedge tip serves as indefinitely thin plasmonic waveguide "core". On the other hand, a metal wedge is comprised of two metal surfaces, each serves as a single surface SPP mode waveguide, which are coupled at their merging point. These two distinct modes of propagation supported by the wedge configuration are manifested by the two dispersion curves. The propagation associated with wedge tip 1D discontinuity is relegated to the upper branch, while the lower curve is more of the two coupled surface modes.

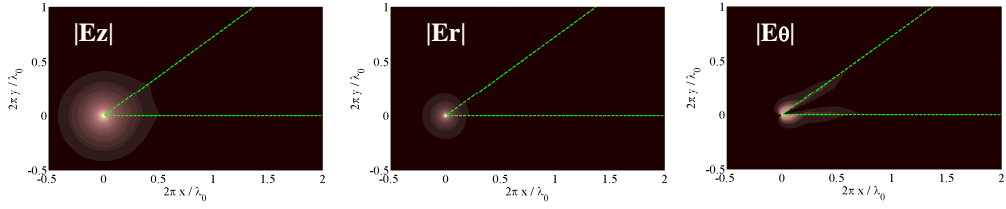

Fig. 2. Electric field components (absolute value, A.U.) at pt. I ($\beta=6k_0; \omega=0.74\omega_p$).

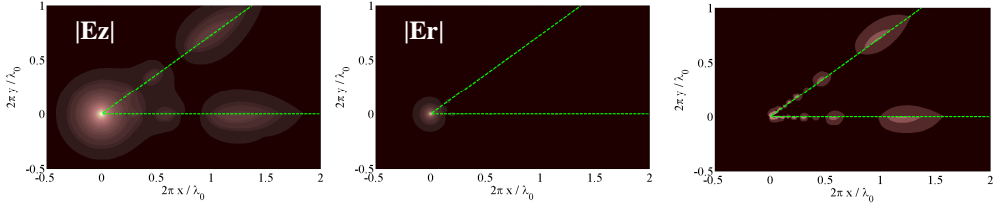

Fig. 3. Electric field components (absolute value, A.U.) at pt. II ($\beta=6k_0; \omega=0.69\omega_p$).

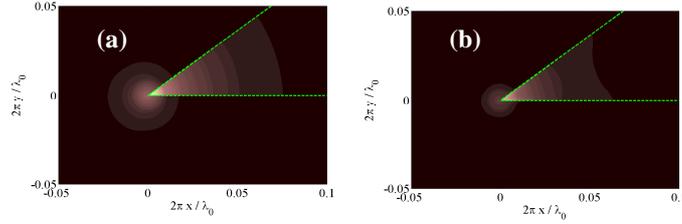

Fig. 4. Tangential Pointing vector (Sz) (absolute value, A.U.) at: (a) pt. II, (b) pt. I

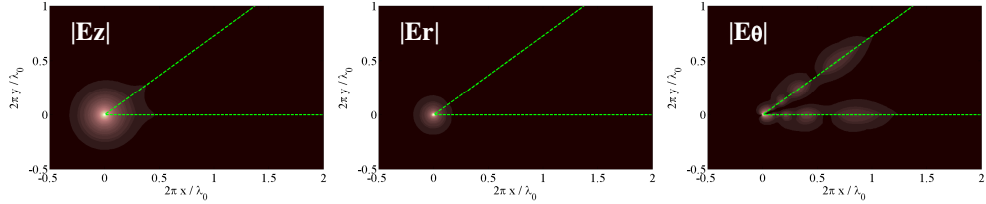

Fig. 5. Electric field components (absolute value, A.U.) at pt. III ($\beta=8k_0; \omega=0.7\omega_p$).

Similarly to slab modes, the field associated with the lower curve (depicted in Fig. 3 for pt. II) has both tangential component ($E_z$) and component normal to the interface ($E_\theta$). For the normal component the two surface modes coupled at the tip are evident, being located on the interfaces. However, the dominant component (in terms of maximal field amplitude) is actually $E_r$, which is the transversal parallel component. This component, which is mainly located around the tip, is absent for slab waveguide, and emerges from the hybrid configuration. Since the field related to the upper branch is mainly guided by the wedge tip (rather than the wedge faces) its field components are exhibiting reduced azimuthal dependence, as illustrated in Fig. 2 for pt. I.

This narrative elucidates the distinctive shape of dispersion curves. For substantial values of $\beta$ the significant structural details for the plasmon waves propagating at each point on an interface are the local neighborhood, resulting in dispersion curve similar to that of a single surface SPP. However SPPs localized on the two wedge facets are exceedingly coupled in the vicinity of the wedge point, which splits the dispersion curve according to the mode symmetry. For $E_z^e$ the dispersion curve is pushed to lower frequency in respect to the single surface SPP curve. This symmetry, of even tangential E-field, is related to the odd potential mode of metal slab (equivalent to odd transversal H-field) [13]. For low $\beta$-values the localization notion is weakening, and the lower branch resembles the upper one for extremely low $\beta$-values. The two different asymptotic behaviors of the lower branch results in the minimal frequency point on the dispersion curve at $\beta=5k_0$.

Comparing of the upper branch of Fig 1(b) with the electrostatic analysis [4], is exhibiting a qualitative similarity for large $\beta$ values: modal frequency is above the asymptotic single surface SPP frequency and decreasing monotonically with $\beta$. This is related to the mode localized on the 1D discontinuity.

Equipped with the dispersion characteristics, sensitivity to structural parameters can be conjectured. As the surrounding dielectric constant increases, the dispersion curves are pulled down, as $\omega_{spp}$ decreases. The cutoff frequency should increase with the wedge angle, as the lower dispersion branch may diverge less from the single surface SPP curve, due to reduced coupling between the modes near the wedge point. A cutoff angle is thus expected at each given wavelength, above which a solution does not exist. This resembles the cutoff angle that was observed in some specific FDTD simulations for different symmetry modes ($E^eH^e$) [7] and also for the complementary structure of V-grooved channel polaritons [10]. It is noteworthy that the modes of the symmetry reported here are have higher effective index (i.e., enhanced plasmonic effect) and tighter confinement, than those reported in [7].

## 4. Conclusion

We analyzed in detail nano-scale modes propagating on a metal waveguide with a wedge cross-section. Two branches are resolved, one is related to the wedge point localized mode (wire like) and the second to the waves propagating on the interfaces crossing at the wedge point. This notion is used to explain the curve characteristics, as well to validate with previous asymptotic studies and observations. The cross-section of the modes is of a substantial sub-wavelength dimension, with most of the guided power propagating within a cross section with each dimension ~1% of the wavelength. Including the metal losses, especially for frequencies near $\omega_P$, yields rapidly attenuated modes. Gain-assisted propagation [14] may mitigate this difficulty.

We acknowledge the Israeli Ministry of Science for support.

**APPENDIX**

The boundary conditions coefficients:

$$f_1 = \frac{\varepsilon_D}{|\varepsilon_M|}\frac{q_M^2}{q_D^2}; \quad f_2 = \frac{\omega\mu\beta(|\varepsilon_M|+\varepsilon_D)}{q_D^2|\varepsilon_M|}; \quad f_3 = \frac{\omega\varepsilon_0\beta(|\varepsilon_M|+\varepsilon_D)}{q_D^2}; \quad f_4 = \frac{q_M^2}{q_D^2}; \tag{A-1}$$

Expressions for G and H functions:

$$G(\xi_1,\xi_2;\tau) = \frac{\pi^2 \cos(\xi_1 \ln(\tau))}{2\xi_1 \sinh(\pi\xi_1)}\delta(\xi_1-\xi_2) + \frac{\pi^2}{16}\left[\cosh(\pi\xi_1)-\cosh(\pi\xi_2)\right]^{-1} g(\xi_1,\xi_2;\tau)$$

$$g(\xi_1,\xi_2;\tau) = \left[\tau^2-1\right]\tau^{-i\xi_1} {}_2\mathbb{F}_1\left(1-i\frac{\xi_1+\xi_2}{2}, 1-i\frac{\xi_1-\xi_2}{2}; 2; 1-\tau^2\right) \tag{A-2}$$

$$H(\xi_1,\xi_2) = \frac{\pi^2}{2}\left[\cosh(\pi\xi_1)-\cosh(\pi\xi_2)\right]^{-1}; \quad G(\xi,\xi;\tau)=0; \quad H(\xi,\xi)=0;$$

${}_2F_1$ denotes the Hypergeometric function.